\documentclass[twocolumn,pra,superscriptaddress]{revtex4}
\usepackage[T1]{fontenc}
\begin{document}
\title{Theoretical Falsification of Leggett Conjecture for Two Qubits or One Ququat}
\author{Marcin Wie\'sniak}
\affiliation{Centre for Quantum Technologies, National University of Singapore, 3 Science Drive 2, Singapore 117543, Singapore}\affiliation{Institute of Theoretical Physics and Astrophysics, University of Gda\'nsk,\\ ul. Wita Stwosza 57, PL-80-952 Gda\'nsk, Poland}
\begin{abstract}
Recently, some attention has been paid to falsifying the Leggett model, in which global probabilities characterizing a quantum state are represented by a combination of factorisable distributions. This idea was even verified in experiments, and generalized for larger subsystems, but thus far not in terms of the size of a subsystem. In this communication we show an inequality to reject the Leggett description for a subsystem of two qubits. We also point out some other interpretations of the derived expression.
\end{abstract}
\maketitle
\section{Introduction}
The unavoidable randomness of Quantum Mechanics (QM) has always been a fact difficult to accept. The debate was initialized by the letter of Einstein, Podolsky, and Rosen \cite{epr}, in which they stated the following alternative: either QM cannot be considered complete as it is unable to predict with certainty, say, the position and the momentum of a particle, or these two measurements are not compatible as two elements of the same reality. The incompleteness of QM, which was the conclusion of \cite{epr} would open a possibility to introduce additional, hidden variables (HV). These parameters would deterministically describe what we percept as randomness. Later, Bohm \cite{Bohm1,Bohm2} has introduced his interpretation of QM, which allowed hidden variables. Indeed, the Bohmian hidden variable model cannot be falsified, as it allows to assign any result to any measurement in a given place and moment  of time. An important step was made in 1964, when Bell \cite{Bell} introduced his inequality to falsify the hypothesis of the existence of local hidden variables. This theorem has been generalized (see e.g. \cite{mermin}) and experimentally trailed (starting with \cite{aspect}) in many various attempts. Another important step toward the rejection of HV modes was the Kochen-Specker theorem \cite{KS}. It stated a fundamental contradiction between hidden parameters and non-commutative algebra of operators (observables).

Recently falsified Leggett model \cite{leggett} is in some sense opposite to the local hidden variable model \cite{epr}. In the latter, we try to reconstruct observed correlations under the assumption that each particle carries a local and predetermined set of orders, which result to yield under any possible measurement. The Bell theorem \cite{Bell} shows our failure in this attempt, and hence the assumption of local realism must be rejected. On contrary, the Leggett model can be described as following: correlations observed in the composite quantum system are true, but there exists a non-local hidden variable $\lambda$, which defines states of subsystems as some pure states. The integration over the set of $\lambda$ with some distribution $\rho(\lambda)$ reconstructs the marginal statistics observed in experiments. Technically the Leggett model states that quantum correlations can be modeled by some integer over statistics of product states:
\begin{eqnarray}
\label{idea1}
&\lambda\leftrightarrow|\vec{u}\rangle^{[1]}|\vec{v}\rangle^{[2]},\\
\label{idea2}
&P_{QM}(a,b|\vec{a},\vec{b})=\int d\lambda \rho(\lambda)P_\lambda (a,b|\vec{a},\vec{b}).
\end{eqnarray}
Here, $a=\pm 1$ and $b=\pm 1$ are results obtained in measurements $A=\vec{a}\cdot\vec{\sigma}^{[1]}$ and $B=\vec{b}\cdot\vec{\sigma}^{[2]}$, conducted on the first and the second qubit, respectively. The marginal probabilities are
\begin{eqnarray}
\label{margin1}
&P_\lambda(a|\vec{a})=\frac{1}{2}(1+ a\vec{u}\cdot\vec{a}),\\
\label{margin2}
&P_\lambda(b|\vec{b})=\frac{1}{2}(1+ b\vec{v}\cdot\vec{b}),
\end{eqnarray}
and since for every value of $\lambda$ the state is a product, one has $P_\lambda(a,b|\vec{a},\vec{b})=P_\lambda(a|\vec{a})P_\lambda(b|\vec{b})$.

While the Bell theorem is falsified by showing the incompatibility of the correlations, the Leggett model is rejected because this description of subsystems violates the positivity of the joint probability. The inequality is based on the fact that for a given $\lambda$ the probability distribution must be physical, satisfying
\begin{equation}
\label{idea3}
-1+|\overline{A}_\lambda-\overline{B}_\lambda|\leq\overline{AB}_\lambda\leq 1-|\overline{A}_\lambda+\overline{B}_\lambda|,
\end{equation}
Where $\bar{\cdot}_\lambda$ denotes the average for a given $\lambda$.
This contradiction turned out to be a successful method of falsifying the non-local model, both in theory and in experiment \cite{simon,vienna2,scarani,scarani2}. The inequalities tested in those experiments, however, allowed to reject only the Leggett-like description of one qubit in the context of a two-qubit experiment. In a more recent report \cite{mpm}, the experimental context--the number of entangled qubits--has been extended to arbitrary $N$, allowing to reject theories with
\begin{equation}
\lambda\leftrightarrow|\vec{v}^{[1]}\rangle^{[1]}|\vec{v}^{[2]}\rangle^{[2]}|\vec{v}^{[3]}\rangle^{[3]}... \quad .
\end{equation}
Still, however, the subsystem under consideration was composed out of only one qubit. Is it possible to reject more general Leggett models? In this contribution we show present a method to exclude a similar description for a subsystem of two qubits in a four-qubit experiment. This time the hidden parameter would be associated with
\begin{equation}
\label{idea4}
\lambda\leftrightarrow|\psi_1\rangle^{[1,2]}|\psi_1\rangle^{[3,4]}.
\end{equation}

\section{Leggett inequality for subsystem of one qubit}
Before we present the new inequality, let us recall the four-qubit expression derived in \cite{mpm}. Let $A,B,C,D$ denote dichotomic measurments with outcomes $\pm 1$, conducted on each of four particles. The probability of obtaining results $a,b,c,d$ is 
\begin{eqnarray}
&P_\lambda(a,b,c,d)\nonumber\\
=&\frac{1}{16}(abcd\overline{ABCD}_\lambda\nonumber\\
+&abc\overline{ABC}_\lambda+abd\overline{ABD}_\lambda+acd\overline{ACD}_\lambda+bcd\overline{BCD}_\lambda\nonumber\\
+&ab\overline{AB}_\lambda+ac\overline{AC}_\lambda+ad\overline{AD}_\lambda\nonumber\\
+&bc\overline{BC}_\lambda+bd\overline{BD}_\lambda+cd\overline{CD}_\lambda\nonumber\\	
+&a\overline{A}_\lambda+b\overline{B}_\lambda+c\overline{C}_\lambda+d\overline{D}_\lambda+1)\nonumber\\\geq &0.
\end{eqnarray}
For $P_\lambda(+,+,+,+)$, $P_\lambda(+,+,-,-)$, $P_\lambda(-,-,+,+)$, and $P_\lambda(-,-,-,-)$, the positivity condition can be jointly written as
\begin{eqnarray}
&\overline{ABCD}_\lambda+\overline{AB}_\lambda+\overline{CD}_\lambda&\nonumber\\+&s(\overline{A}_\lambda+\overline{BCD}_\lambda+\overline{B}_\lambda+\overline{ACD}_\lambda)&\nonumber\\
\pm&|\overline{AC}_\lambda+\overline{BD}_\lambda+\overline{AD}_\lambda+\overline{BC}_\lambda&\nonumber\\
	+&s(\overline{C}_\lambda+\overline{ABD}_\lambda+\overline{D}_\lambda+\overline{ABC}_\lambda)|+1&\nonumber\\
\geq &0
\end{eqnarray}
($s=\pm 1$). Dropping the second modulo we get
\begin{equation}
\label{eval2}
\overline{ABCD}_\lambda+\overline{AB}_\lambda+\overline{CD}_\lambda-|\overline{A}_\lambda+\overline{BCD}_\lambda+\overline{B}_\lambda+\overline{ACD}_\lambda|+1\geq 0.
\end{equation}
A similar derivation shall be now done with $P_\lambda(+,-,+,-)$, $P_\lambda(+,-,-,+)$, $P_\lambda(-,+,+,-)$, and
$P_\lambda(-,+,-,+)$. In that process we obtain
\begin{equation}
\label{eval2a}
\overline{ABCD}_\lambda-\overline{AB}_\lambda-\overline{CD}_\lambda-|\overline{A}_\lambda+\overline{BCD}_\lambda-\overline{B}_\lambda-\overline{ACD}_\lambda|+1\geq 0.
\end{equation}
Let us now add (\ref{eval2}) and (\ref{eval2a}) sidewise and use the triangle inequality $|A\pm B|\leq |A|+|B|$:
\begin{equation}
\label{eval3}
\overline{ABCD}_\lambda-|\overline{A}_\lambda+\overline{BCD}_\lambda|+1\geq 0.
\end{equation}
Subsequently, we replace $A$ with some $A'$ and again employ the triangle inequality:
\begin{equation}
\label{eval4}
\overline{ABCD}_\lambda+\overline{A'BCD}_\lambda-|\overline{A}_\lambda-\overline{A'}_\lambda|+2\geq 0.
\end{equation}

To falsify the Leggett hypothesis, we follow \cite{scarani2} and use the fact that for normalized vector $\vec{v} (\vec{v}^2=1)$, one has that the taxi metric $\sum_{i=1}^3|\vec{e}_i\cdot\vec{v}|\geq 1$, with $\vec{e}_1,\vec{e}_2,\vec{e}_3$ are normalized and define the three axes. Let us now introduce three sets of observables to be able to use the taxi metric. The first set is given by
\begin{eqnarray}
&\vec{a}_1=\cos 2\alpha\vec{e}_1+\sin 2\alpha\vec{e}_2,&\nonumber\\
&\vec{a}_1'=\cos 2\alpha\vec{e}_1-\sin 2\alpha\vec{e}_2,&\nonumber\\
&-\vec{b}_1=\vec{c}_1=\vec{d}_1=\vec{e}_1,&\nonumber
\end{eqnarray}
and the other two are obtained by a cyclic permutation of $\vec{e}_1,\vec{e}_2$, and $\vec{e}_3$. The factor in front of $\alpha$ will turn out to be useful later. The sum over the three sets gives
\begin{eqnarray}
\label{leggett1}
&\sum_{i=1}^3\overline{A_iB_iC_iD_i}_\lambda+\overline{A_i'B_iC_iD_i}_\lambda\nonumber\\
\geq&-6+\sum_{i=1}^3|\overline{A_i}_\lambda-\overline{A_i'}_\lambda|.&
\end{eqnarray}
After the integration over $\lambda$ the averages in the modulo on the right-hand had side are, $2|\sin 2\alpha|\int d\lambda\rho(\lambda)|\vec{v}^{[1]}(\lambda)\cdot\vec{v}_i|$, with $\vec{v}_i$ being $\vec{e}_2$, $\vec{e}_3$, and $\vec{e}_1$, respectively, for $i=1,2,3$, and $\vec{v}^{[1]}(\lambda)$ denoting the Bloch vector of the first qubit for $\lambda$. Using the supremacy of the taxi metric, we can bound the modulo from below:
\begin{equation}
\sum_{i=1}^3|\overline{A_i}_\lambda-\overline{A_i'}_\lambda|\geq 2|\sin 2\alpha|.
\end{equation}
Putting the quantum mechanical values for the GHZ state, $|GHZ\rangle=\frac{1}{\sqrt{2}}(|0000\rangle+|1111\rangle)$, we observe the violation of (\ref{leggett1}) for $0<|\alpha|<0.10\pi$. This shows us that the one-particle Leggett model is not possible in a four-qubit experiment.

\section{Subsystem of two qubits}
Let us now consider the two-qubit Leggett-like description. We come back to the inequalities (\ref{eval2}) and (\ref{eval2a}), which together state that
\begin{equation}
\label{eval5}
\overline{ABCD}_\lambda-|\overline{AB}_\lambda+\overline{CD}_\lambda|+1\geq 0.
\end{equation}
We again introduce alternative observables $A'$ and $B'$ and applying the triangle inequality twice obtain
\begin{equation}
\label{eval6a}
\overline{(A+A')(B+B')CD}_\lambda-|\overline{(A+A')(B-B')}_\lambda|+4\geq 0
\end{equation}
\begin{equation}
\label{eval6b}
\overline{(A+A')(B+B')CD}_\lambda-|\overline{(A-A')(B+B')}_\lambda|+4\geq 0.
\end{equation}
We now propose the following sets of vectors defining observables:

\begin{tabular}{|c|cccccc|}
\hline
&&&&&&\\
$i$&$\vec{a}_i$&$\vec{a}'_i$&$\vec{b}_i$&$\vec{b}'_i$&$\vec{c}_i$&$\vec{d}_i$\\
\hline
4&$m\vec{e}_1+n\vec{e}_2$&$m\vec{e}_1-n\vec{e}_2$&$m\vec{e}_1+n\vec{e}_2$&$m\vec{e}_1-n\vec{e}_2$&$-\vec{e}_1$&$\vec{e}_1$\\
5&$m\vec{e}_1+n\vec{e}_3$&$m\vec{e}_1-n\vec{e}_3$&$m\vec{e}_1+n\vec{e}_3$&$m\vec{e}_1-n\vec{e}_3$&$\vec{e}_2$&$\vec{e}_2$\\
6&$m\vec{e}_1+n\vec{e}_2$&$m\vec{e}_1-n\vec{e}_2$&$m\vec{e}_2+n\vec{e}_1$&$m\vec{e}_2-n\vec{e}_1$&$\vec{e}_2$&$\vec{e}_1$\\
7&$m\vec{e}_2+n\vec{e}_3$&$m\vec{e}_2-n\vec{e}_3$&$m\vec{e}_2+n\vec{e}_3$&$m\vec{e}_2-n\vec{e}_3$&$-\vec{e}_2$&$\vec{e}_2$\\
\hline
\end{tabular}
with $m=\cos\alpha$ and $n=\sin 2\alpha$. From (\ref{eval6a}) and (\ref{eval6b}) we construct
\begin{eqnarray}
\label{eval7}
&2\sum_{i=4}^{7}\overline{(A_i+A'_i)(B_i+B'_i)C_iD_i}_\lambda\nonumber\\
\geq&-64+\sum_{i=4}^7(|\overline{(A_i+A'_i)(B_i-B'_i)}_\lambda|\nonumber\\
+&|\overline{(A_i-A'_i)(B_i+B'_i)}_\lambda|).
\end{eqnarray}

At the right-hand side, all moduli will have a common factor $4|\sin\alpha||\cos\alpha|=2|\sin 2\alpha|$. It is also important that each of eight moduli depends on a different pair of the unit vectors $\vec{e}_i^{[1]}\cdot\vec{e}_j^{[2]}$. The only pair that never appears is $\vec{e}_3^{[1]}\cdot\vec{e}_3^{[2]}$. This is precisely because the GHZ state possesses this correlation, $T_{3300}=1$.

 Is ineq. (\ref{eval7}) sufficient to falsify the Leggett model for two qubits? We know that for pure states of two qubits $1\leq\sum_{i,j=1}^3T_{ij}^2\leq 3$. The minimum is reached for product states, and the maximum refers to the maximally entangled ones. Considering the worst case scenario, all states assigned by $\lambda$ could be products with only non-zero correlation $T_{33}$. Inequality (\ref{eval7}) does not exclude this possibility by itself. For this reason we combine it with (\ref{leggett1}) and obtain
\begin{eqnarray}
\label{eval8}
&\sum_{i=1}^3(\overline{(A_i+A'_i)B_iC_iD_i}_\lambda+\overline{A_i(B_i+B'_i)C_iD_i}_\lambda^\triangle)\nonumber\\
+&2\sum_{i=4}^{7}\overline{(A_i+A'_i)(B_i+B'_i)C_iD_i}_\lambda\nonumber\\
\geq& -76+\sum_{i=1}^3\left(|\overline{A_i-A'_i}_\lambda|+|\overline{B_i-B'_i}_\lambda|^\triangle\right)\nonumber\\
+&\sum_{i=4}^7\left(|\overline{(A_i+A'_i)(B_i-B'_i)}_\lambda|\right.\nonumber\\
+&\left.|\overline{(A_i-A'_i)(B_i+B'_i)}_\lambda|\right),\nonumber\\
\end{eqnarray}
where the triangle denotes that the assignment of settings is interchanged for the first and second qubit.

Inequality (\ref{eval8}) must be true for all values of the Leggett's hidden parameter, thus the right-hand side thereof is integrated over $\lambda$:
\begin{eqnarray}
\label{eval8a}
&\sum_{i=1}^3\left(\overline{(A_i+A'_i)B_iC_iD_i}+\overline{A_i(B_i+B'_i)C_iD_i}^\triangle\right)\nonumber\\
+&2\sum_{i=4}^{7}\overline{(A_i+A'_i)(B_i+B'_i)C_iD_i}\nonumber\\
\geq& -76+\int d\lambda\rho(\lambda)\left(\sum_{i=1}^3(|\overline{A_i-A'_i}_\lambda|+|\overline{B_i-B'_i}_\lambda|^\triangle)\right.\nonumber\\
+&\sum_{i=4}^7(|\overline{(A_i+A'_i)(B_i-B'_i)}_\lambda|\nonumber\\
+&\left.|\overline{(A_i-A'_i)(B_i+B'_i)}_\lambda|)\right),\nonumber\\
\end{eqnarray}
 
Let us now adopt the formalism of the correlation tensor. Its elements are taken as 
\begin{equation}
T_{ijkl}=\langle \psi|\sigma_i^{[1]}\sigma_j^{[2]}\sigma_k^{[3]}\sigma_l^{[4]}|\psi\rangle.
\end{equation}
 Now we can use the fact that for any pure two-qubit state $\sum_{i,j=0}^3T_{ij00}^2=4$. $T_{0000}=1$ trivially expresses the normalization condition. Given that $|T_{3300}|\leq 1$, we still have $\sum_{i,j=0}^3|T_{ij00}|-|T_{3300}|-T_{0000}\geq 2$. This is because every element of the correlation tensor is not larger in modulo than 1; the taxi metric is larger than the square root of the Euclidean metric. Applying this bound and integrating over $\lambda$ one gets
\begin{eqnarray}
\label{leggett2}
&\sum_{i=1}^3(\overline{(A_i+A'_i)B_iC_iD_i}+\overline{A_i(B_i+B'_i)C_iD_i}^\triangle)\nonumber\\
+&2\sum_{i=4}^{7}\overline{(A_i+A'_i)(B_i+B'_i)C_iD_i}\nonumber\\
\geq& -76+4|\sin 2\alpha|,\nonumber\\
\end{eqnarray}

The left-hand side of (\ref{leggett2}) shall be again computed for the GHZ state. Let us characterize the state in terms of the four-qubit correlation tensor:
\begin{eqnarray}
\label{corten}
1=&T_{3333}=T_{1111}=T_{2222}\nonumber\\
=&-T_{1122}=-T_{1212}=-T_{2112}\nonumber\\
=&-T_{1221}=-T_{2121}=-T_{2211},
\end{eqnarray}
and the rest of four-qubit correlations vanishes. With these correlation tensor elements the quantum mechanical averages read
\begin{eqnarray}
\label{qmav}
&\overline{A_iB_iC_iD_i}=\overline{A'_iB_iC_iD_i}=\overline{A'_iB_iC_iD_i}^\triangle\nonumber\\
=&\overline{A'_iB'_iC_iD_i}^\triangle=-\cos 2\alpha\quad(i=1,2,3)\nonumber\\
&\overline{A_4B'_4C_4D_4}=\overline{A'_4B_4C_4D_4}\nonumber\\
=&\overline{A_6B_6C_6D_6}=\overline{A'_6B'_6C_6D_6}=-1,\nonumber\\
&\overline{A_4B_4C_4D_4}=\overline{A'_4B'_4C_4D_4}\nonumber\\
=&\overline{A'_6B_6C_6D_6}=\overline{A_6B'_6C_6D_6}=-\cos 2\alpha,\nonumber
\end{eqnarray}
and other values we need are equal to $-\cos_2\alpha$. Thus the left-hand side adds up to $-16(1+\cos 2\alpha+2\cos^2 \alpha)-12\cos 2\alpha=-32-44\cos 2\alpha$.

\section{Summary}
The inequality is violated for $0\leq |\alpha|\leq 0.0289\pi$. The maximal violation reads $0.181444$, which translates into the maximal allowed admixture of the white noise $0.238\%$. It will probably become within experimental possibilities not before long, but the result has also some interesting theoretical consequences. First, it clearly shows that the Leggett-like model for subsystems of two qubits is not possible. In more practical terms, this means that there is, in general, no necomposition of the four-qubit statistics in the sense of (\ref{idea4}). This product non-decomposability is expected, but here it has been demonstrated explicitly. This is in agreement with obsevations from Ref. \cite{entent}. Krenn and Zeilinger have noticed that in a many-qubit GHZ state violation of a Bell inequality for two qubits can be observed only conditionally on measurements on other qubits (the dscusses example was of $N=3$). 

 We may also give another interpretation of (\ref{eval8}). What we have really shown is that when a four-qubit state possesses the correlations used in the inequality then the only non-vanishing elements of the correlation tensor of a two-qubit subsystem state are $T_{3300}$, $T_{3030}$, $T_{3003}$, $T_{0330}$, $T_{0303}$, and $T_{0033}$. Similarly, from the original two qubit-Leggett inequality one can infere that if the state has one perfect correlation, e.g. $T_{33}=1$, Bloch vectors of individual qubits can have only $z$ components. If the two qubit state has three (or even two) perfect correlations, the Bloch vectors necessarily vanish. This could be seen as a complementarity relation: the weaker are $N$-partite correlations in the state (the peak of the cosine-like function on the left-hand side), the more information is allowed in the reduced state (the factor in front of $|\sin 2\alpha|$ at the right-hand side), so that the two curves do not cross foor sufficiently weak correlations. 

The Reader can easily extend ineq. (\ref{eval8}) to more qubits ``separated'' in the form of $(2|N-2)$. This generalization would once more involve the GHZ states and their perfect correlations in the $xy$ planes. If $N$ is odd, all $\vec{e}_3$s in the third set of settings shall be replaced with $\vec{e}_1$s.    

For the independent derivation of Leggett-type inequalities, please refer to Ref. \cite{chen}


\section{Acknowledgements}
The Author wishes to thank V. Scarani for useful discussions. This work was supported by the National Research  
Foundation and Ministry of Education, Singapore and by a MNiSW Grant IdP2011 00361 (Ideas Plus).

\end{document}